\documentclass[twocolumn,showpacs,preprintnumbers,aps,amsmath,amssymb]{revtex4}

\usepackage{graphicx}
\usepackage{dcolumn}
\usepackage{bm}
\usepackage{epsfig}

\begin{document}


\title{Pragmatical access to the viscous flow of undercooled liquids}

\author{U. Buchenau}
 \email{buchenau-juelich@t-online.de}
\affiliation{%
J\"ulich Center for Neutron Science, Forschungszentrum J\"ulich\\
Postfach 1913, D--52425 J\"ulich, Federal Republic of Germany
}%

\date{May 11, 2017}

\begin{abstract}
The paper derives a relation for the viscosity of undercooled liquids on the basis of the pragmatical model concept of Eshelby relaxations with a finite lifetime. From accurate shear relaxation data in the literature, one finds that slightly less than half of the internal stresses relax directly via single Eshelby relaxations; the larger part dissolves at the terminal lifetime, which is a combined effect of many Eshelby relaxations. 
\end{abstract}

\maketitle

The shear response of undercooled liquids is much better measured \cite{plazek-magill,plazek-bero,mills,schro,donth1,donth2,niss1,maggi,pcnew,tina} than it is theoretically understood \cite{cavagna,bb}. 

The problem is intimately related to the existence of a terminal lifetime $\tau_c$. The existence of such a lifetime, after which the final relaxation is exponential, has been shown in aging experiments \cite{tina2010}. It is experimentally well established \cite{tina2010,bo} that this lifetime is markedly longer than the Maxwell time $\tau_M=\eta/G$.

At shorter times, the shear relaxation proceeds with the Kohlrausch time dependence $t^\beta$, with a Kohlrausch $\beta$ close to 1/2. The relaxation processes behind this Kohlrausch behavior must be back-and-forth jumps, because one finds a recoverable compliance $J_0$ which is more than a factor of two larger \cite{plazek-magill,plazek-bero} than the elastic recoverable compliance $1/G$ alone.

Thus there must be a crossover from back-and-forth jumps at short times to no-return jumps at the terminal lifetime $\tau_c$, for which there is not yet a theory \cite{cavagna,bb}.

A first attempt to attack this problem is the pragmatical model of Eshelby relaxations with a finite lifetime \cite{visc}. The present paper develops this model further and derives a relation for the ratio $\tau_c/\tau_M$  between terminal and Maxwell time.

With this relation, accurate shear relaxation data allow to determine the fraction $f_c$ of the internal stresses which are not relaxed by any single Eshelby relaxation, but dissolve in the collective terminal process. The recognition of the fraction $f_c$ is important for the description of dielectric and other non-shear relaxation data, because it increases their viscous components. In particular, $f_c$ is crucial for the understanding of nonlinear dielectric data \cite{nonlin}.

The pragmatical model \cite{visc} has six parameters: The infinite frequency shear modulus $G$, the viscosity $\eta$, the recoverable compliance $J_0$, the Kohlrausch $\beta$, the structural lifetime $\tau_c$, and, finally, the full width $W_r$ at half maximum of of the distribution of local structural lifetimes on a logarithmic scale. 

Since measured data are only rarely accurate enough to fix six parameters, this implies an undesirable arbitrariness in the results of the fitting procedure \cite{visc}.

On the other hand, none of the six parameters is avoidable. In the case of the shear modulus $G$, the viscosity $\eta$ and the Kohlrausch $\beta$, this is generally agreed. The recoverable compliance $J_0$, though no longer in the focus of the attention of the community, is an acknowledged textbook \cite{ferry} material constant. 

Finally, since one has dynamical heterogeneity in undercooled liquids \cite{hetero}, one must also reckon with a distribution of local structural lifetimes.

In the pragmatical model \cite{visc}, each double-well potential stands for a local cooperative rearrangement. The rearranging region changes its volume and its shape in the rearrangement, coupling to the surrounding elastic matrix according to the Eshelby theory \cite{eshelby}.

In the Eshelby theory, the structural jump of the central core couples to the stress, not to the strain \cite{eshelby}. This implies that the effects of different structural jumps do not add in the shear modulus, but in the elastic shear compliance \cite{ferry}
\begin{equation}\label{jom}
J(\omega)=\frac{1}{G}+\int_{-\infty}^\infty \frac{L(\tau)}{1+i\omega\tau}d\ln\tau-\frac{i}{\omega\eta},
\end{equation}

The retardation function $L(\tau)$ defines the zero-frequency recoverable compliance $J_0$, the infinite frequency elastic compliance plus the integral over the retardation processes
\begin{equation}\label{j0e}
	J_0=\frac{1}{G}+\int_{-\infty}^\infty L(\tau)d\ln\tau.
\end{equation}

In the absence of a secondary relaxation peak, the density of local cooperative rearrangements in the logarithm $\ln\tau_r$ of their relaxation time is given by
\begin{equation}
	l(\tau_r)=l_0(\tau_r/\tau_c)^\beta,
\end{equation}
in order to reproduce the rise of the shear response with the Kohlrausch $t^\beta$.

A given local rearrangement with relaxation time $\tau_r$ does not exist forever, because the surroundings flow, providing it with a decay time $\tau_c$.

The rates $r=1/\tau_r$ and $r_c=1/\tau_c$ add up to the total decay rate $1/\tau=r+r_c$, so
\begin{equation}\label{tau}
\tau=\frac{1}{r+r_c}=\frac{\tau_r\tau_c}{\tau_r+\tau_c}	
\end{equation}
which is always shorter than both $\tau_r$ and $\tau_c$. 

The total response probability of the rearrangement within its lifetime is given by
\begin{equation}
	\frac{r}{r+r_c}=\frac{\tau_c}{\tau_c+\tau_r}
\end{equation}
(in the first version \cite{visc}, this was mistakenly taken to be the retardational response probability).

This probability divides into a probability for viscous response
\begin{equation}
	\frac{rr_c}{(r+r_c)^2}=\frac{\tau_r\tau_c}{(\tau_c+\tau_r)^2}
\end{equation}
and a probability for retardation response 
\begin{equation}
	\frac{r^2}{(r+r_c)^2}=\frac{\tau_c^2}{(\tau_c+\tau_r)^2}.
\end{equation}

Integrating both components over $\tau_r$, one finds the total integral $I_\beta$ (in reasonable approximation) 
\begin{equation}\label{ibet}
	I_\beta=l_0\pi(1+13.27(\beta-1/2)^2+135(\beta-1/2)^4)
\end{equation}
with the exact retardation fraction
\begin{equation}\label{fr}
	f_r=1-\beta.
\end{equation}

The resulting retardation function is 
\begin{equation}\label{norm}
	L(\tau)d\ln{\tau}=\frac{\tau_c^2}{(\tau_r+\tau_c)^2}\frac{l(\tau_r)}{G}d\ln{\tau_r},
\end{equation}
which evaluates to
\begin{equation}
	L(\tau)=\frac{l_0}{G}\frac{\tau_c-\tau}{\tau_c}\left(\frac{\tau}{\tau_c-\tau}\right)^\beta.
\end{equation}

A fraction $f_r$ of the relaxation shear response goes into the retardation $J_0-1/G$ and a fraction $1-f_r$ is viscous response at the average relaxation time $\tau_c$.

This implies a contribution to the viscosity from the no-return processes
\begin{equation}\label{etas}
	\frac{\tau_c}{\eta}\mid_1=\frac{\beta}{1-\beta}(J_0-1/G)=\frac{1-f_r}{f_r}(J_0-1/G).
\end{equation}
The formulation in terms of $f_r$ is needed for cases with excess wings and secondary relaxations.

Note that this part of the viscosity is not a collective effect; it is due to individual Eshelby jumps which are not followed by a back-jump.

The collective effect of all Eshelby jumps leads to the finite structural renewal lifetime $\tau_c$, the decay time for all internal stresses, even in those shear stress direction to which the local rearrangement does not couple. Defining $f_c$ as the fraction of internal stresses released by the collective effect, only the fraction $1-f_c$ is left for the direct effect of the Eshelby jumps. This in turn implies a fraction $(1-f_c)f_r$ responsible for the retardation $J_0-1/G$.

With this normalization, the collective contribution to the viscous flow is
\begin{equation}\label{etac}
	\frac{\tau_c}{\eta}\mid_2=\frac{f_c}{(1-f_c)f_r}(J_0-1/G).
\end{equation}

\begin{figure}
\hspace{-0cm} \vspace{0cm} \epsfig{file=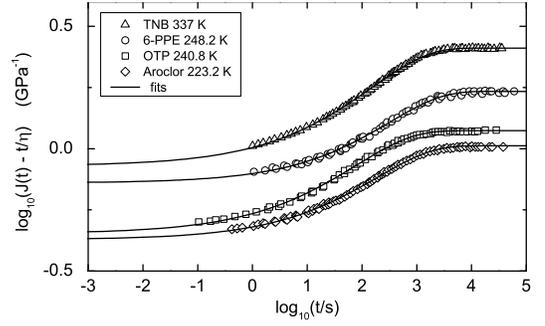,width=7 cm,angle=0} \vspace{0cm} \caption{Fit of recoverable compliance data \cite{plazek-magill,plazek-bero} in terms of the pragmatical model.}
\end{figure}

There is a third contribution: the relaxing region is surrounded by a flowing liquid. The flow of the surroundings induces additional jumps in the relaxing core, which have to be added to the inverse viscosity (the fluidity).

In order to calculate the third contribution, note first that the decay of the cooperatively rearranging region is a continuous process, induced by many rearrangements in the surroundings. After the time $dt$, the state can be considered to consist of a fraction $dt/\tau_c$ of a new state and a fraction $1-dt/\tau_c$ of the old one. 

In continuous flow, the new state is created in equilibrium with the actual shear state at the given time. This continuous creation process implies a back-lag of the equilibrium shear position $\epsilon_{eq}$ of the state by $\sigma\tau_c/2\eta$.

Thus, if one interrupts the continuous flow by taking the external shear stress away at time zero and shear strain zero, the sample relaxes within a few picoseconds (the adaptation time of the vibrations) to the shear strain $\epsilon(0)=-\sigma/G$. Then, as the local relaxations proceed, the sample relaxes gradually to $\epsilon(\infty)=-\sigma J_0$, where $J_0$ is the recoverable compliance \cite{ferry}.

To fulfill this, the actual average equilibrium shear position $\epsilon_{eq}(0)$ of the local relaxations must be at a negative value
\begin{equation}
	\epsilon_{eq}(0)=-\frac{\sigma(J_0-1/G)}{f_r}
\end{equation}
where the factor $1/f_r$ accounts for the fraction of local relaxations decaying without a back-jump.

Equating the two expressions for $\epsilon_{eq}$, one finds
\begin{equation}
	\frac{\tau_c}{\eta}\mid_3=\frac{2}{f_r}(J_0-1/G)=\frac{2}{1-\beta}(J_0-1/G).
\end{equation}

Adding all three contributions, one gets the full fluidity
\begin{equation}\label{eta}
	\frac{\tau_c}{\tau_M}=\frac{3-f_r}{f_r}(GJ_0-1)+\frac{f_c}{f_r(1-f_c)}(GJ_0-1)
\end{equation}
where $\tau_M$ is the Maxwell time $\eta/G$.

Fitting the pragmatical model parameters $\tau_c$, $\tau_M$, $GJ_0$ and $f_r$ to shear relaxation data, one can determine the ratio $f_c$.

The ratio $f_c$ is most accurately obtained from mechanical recoverable compliance measurements \cite{plazek-magill,plazek-bero}, which determine $G$, $\eta$ and $J_0$ directly. $W_r$ has little influence on the curves, so one has essentially only $\tau_c$ and $\beta$ as free parameters. Fig. 1 shows the fits.

\begin{table}[htbp]
	\centering
		\begin{tabular}{|c|c|c|c|c|c|c|c|c|}
\hline
substance                  & $T$  & $G$   & $GJ_0$ & $\beta$& $W_r$ & $\tau_M$& $\tau_c/\tau_M$ & $f_c$    \\
\hline   
                           & $K$  & $GPa$ &        &        &       &  $s$    &         &          \\
\hline                                                                                           
TNB \cite{plazek-magill}   & 337  & 1.18  & 3.03   &  0.37  &   0.0 & 195.0   & 14.2    &0.67      \\
aroclor \cite{plazek-bero} & 232.8& 2.36  & 2.42   &  0.37  &   0.6 & 236.6   & 12.4    &0.75      \\
OTP \cite{plazek-bero}     & 240.8& 2.22  & 2.64   &  0.37  &   0.9 & 104.7   & 13.2    &0.73      \\
6-PPE \cite{plazek-bero}   & 248.2& 1.38  & 2.37   &  0.39  &   0.3 & 344     & 12.7    &0.78      \\
\hline
DC704 \cite{tina}          & 214  & 1.09  & 2.49   &  0.46  &   1.8 & 0.79    & 11.7    &0.65      \\
DC704 \cite{tina}          & 216  & 1.05  & 2.43   &  0.46  &   2.0 & 0.16    & 11.2    &0.62      \\
DC704 \cite{tina}          & 218  & 1.02  & 2.44   &  0.47  &   2.6 & 0.041   & 13.3    &0.66      \\
\hline
5-PPE \cite{tina}          & 250  & 1.03  & 2.33   &  0.50  &   2.0*& 0.69    & 9.5     &0.53      \\
5-PPE \cite{tina}          & 252.5& 0.98  & 2.39   &  0.48  &   2.0*& 0.13    &10.3     &0.58      \\
5-PPE \cite{tina}          & 255  & 0.95  & 2.45   &  0.47  &   2.0*& 0.022   & 11.0    &0.61      \\
\hline
PC \cite{pcnew}            & 159  & 1.48  & 2.96   &  0.47  &   1.7 & 0.34    & 12.5    &0.54      \\
PC \cite{pcnew}            & 161  & 1.42  & 2.97   &  0.49  &   2.0 & 0.037   & 12.1    &0.50      \\
PC \cite{pcnew}            & 163  & 1.39  & 3.14   &  0.55  &   2.8 & 0.006   & 13.0    &0.39      \\
\hline
PG \cite{maggi}            & 171  & 4.08  & 6.39   &  0.53  &   1.6*& 0.45    & 35.9    &0.48      \\
PG \cite{maggi}            & 174  & 3.85  & 6.68   &  0.54  &   1.6*& 0.088   & 38.9    &0.48      \\
PG \cite{maggi}            & 177  & 3.69  & 6.78   &  0.54  &   1.6*& 0.020   & 41.1    &0.52      \\
\hline

		\end{tabular}
	\caption{Determination of the collective fraction $f_c$ of the viscous shear response by applying eq. (\ref{eta}) to recoverable compliance data (TNB is tri-naphtyl benzene, OTP is ortho-terphenyl, 6-PPE is 6-polyphenylether) and to high accuracy $G(\omega)$-data (DC704 is a vacuum pump oil, 5-PPE is 5-polyphenylether, PC is propylene carbonate and PG is propylene glycol). Values marked with an asterisk are taken from fits of dielectric data. PC has an excess wing $f_{1/6}(\tau_r/\tau_c)^{1/6}$ with $f_{1/6}=0.035$, PG has an excess wing with $f_{1/6}=0.027$ (see text).}
	\label{tab:Comp}
\end{table}

Table I compiles the determinations of the collective viscous flow fraction $f_c$ from recoverable compliance and $G(\omega)$ data. The most accurate results in Table I are the first four recoverable compliance measurements \cite{plazek-magill,plazek-bero}, which fix $f_c$ within a few percent.

\begin{figure}
\hspace{-0cm} \vspace{0cm} \epsfig{file=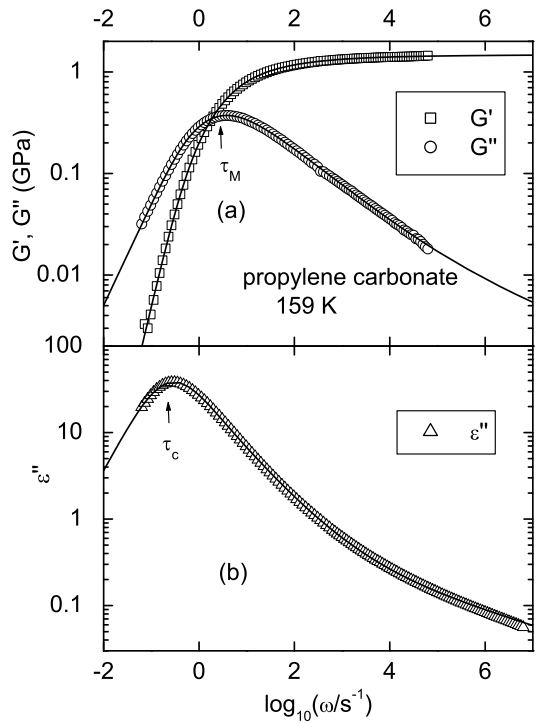,width=7 cm,angle=0} \vspace{0cm} \caption{Fit of (a) $G(\omega)$ data and (b) dielectric relaxation data of propylene carbonate \cite{pcnew} in terms of the pragmatical model.}
\end{figure}

A helpful result from these fits is that the width $W_r$ of the lifetime distribution is comfortably small, of the order of half a decade (remember that a simple Debye relaxation peak has already a width of about a decade). This conclusion is corroborated independently by fits of dielectric data in terms of the pragmatical model, where the viscous component is directly seen at the peak.

$G(\omega)$ data give much less accurate $f_c$-values, because the viscous component overshadows the retardational one at the crossover. Nevertheless, very accurate data like those in DC704 and 5-PPE \cite{tina} determine $f_c$ within an error of about 20 percent. In some of these fits, $W_r$ was taken from fits of dielectric data from the same substance.

\begin{figure}
\hspace{-0cm} \vspace{0cm} \epsfig{file=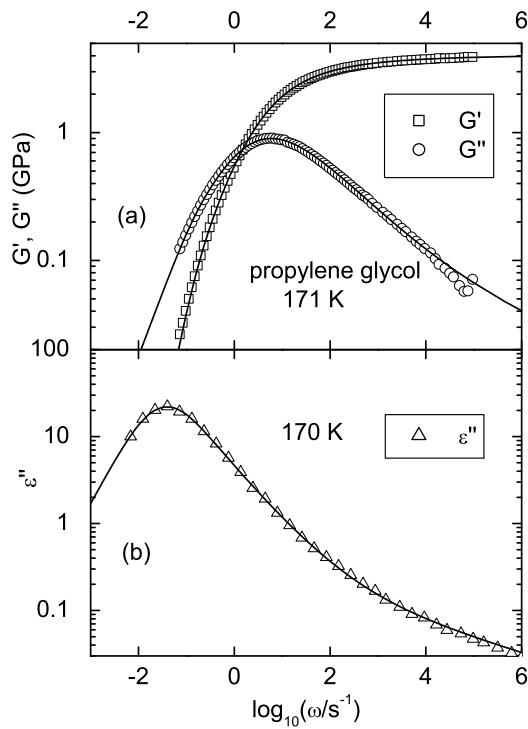,width=7 cm,angle=0} \vspace{0cm} \caption{Fit of (a) $G(\omega)$ data \cite{maggi} and (b) dielectric relaxation data \cite{lunken} of propylene glycol in terms of the pragmatical model.}
\end{figure}

Another new accurate data set exists for propylene carbonate \cite{pcnew}. Like many other strong dielectric substances, propylene carbonate shows an excess wing with a logarithmic slope of about $-1/6$ at high frequencies in its dielectric response (see Fig. 2 (b)). In the new measurement \cite{pcnew}, the authors demonstrated that the same excess wing slope appears also in the shear response.

The excess wing requires the relaxation density
\begin{equation}\label{exc}
	l(\tau_r)=l_0((\tau_r/\tau_c)^\beta+f_{1/6}(\tau_r/\tau_c)^{1/6}
\end{equation}
with the same $f_{1/6}$ of 0.03 in propylene carbonate \cite{pcnew} and glycerol \cite{visc}.

With equs. (\ref{ibet}) and (\ref{fr}), it is straightforward to calculate the retardation fraction $f_r$ for this relaxation density. Inserting $f_r$ and the other fit parameters into eq. (\ref{eta}), one can again determine $f_c$ (see Table I).

With the retardation density $L(\tau)$ obtained from the $l(\tau_r)$ in eq. (\ref{exc}), one can calculate the normalized dielectric susceptibility
\begin{equation}
	\Phi(\omega)=\frac{\epsilon(\omega)-\epsilon_\infty}{\epsilon_s-\epsilon_\infty},
\end{equation}
where $\epsilon_s$ is the static dielectric susceptibility and $\epsilon_\infty$ is its high frequency limit.

To do the calculation, one has to normalize $L(\tau)$ to $f_r(1-f_c)$ and add a viscous component at $\tau_c$ with weight $1-f_r(1-f_c)$, a gaussian distribution of Debye processes with the width $W_r$.

Though the dielectric fits do not determine $f_c$ very accurately, one gets generally smaller $f_c$-values (about 1/3) than the average shear-$f_c$ of 5/9 from Table I. This is understandable: an Eshelby relaxation couples to one of the three electric polarization components, but only to one out of five shear stress components. Thus a factor of 5/3 more shear equilibration is left to the terminal process.

Applying this fit procedure to shear and dielectric data obtained by measuring a propylene carbonate sample from the same charge in the same cryostat \cite{pcnew}, one finds not only the same terminal relaxation time $\tau_c$, but also the same excess wing coefficient $f_{1/6}$ within experimental accuracy. Fig. 2 (a) and (b) shows the shear and dielectric measurements at 159 K.

The same procedure has been followed to evaluate the high accuracy shear data of propylene glycol \cite{maggi} in Fig. 3 (a). The evaluation of dielectric data \cite{lunken} (see Fig. 3 (b)) supplies an excess wing coefficient $f_{1/6}=0.027$, which was taken to be the right one for the fit of the shear data. 

In this case, shear and dielectric data, though measured in different laboratories, show essentially again the same $\tau_c$ for the dielectric data. Since the dielectric relaxation peak coincides with the dynamic heat capacity one \cite{ngai-rendell}, the results support the concept of a terminal relaxation time.  

This is no longer true for DC704. In fact, the correction of the pragmatical model destroys the seemingly perfect agreement between shear and dielectrics reported in the first paper \cite{visc} for DC704. The shear $\tau_c$ of the corrected version is now a factor of 2.4 larger than the dielectric one and lies close to the ones suggested by the peak positions of heat capacity and thermal expansion data \cite{bo} of DC704.

To conclude, the paper derives a relation between the structural lifetime and the Maxwell time, which allows to calculate the collective fraction of the viscous response (the one resulting from the cumulative effect of all Eshelby relaxations together) from accurate shear relaxation data.


\begin{thebibliography}{99}
\bibitem{plazek-magill} D. J. Plazek and J. H. Magill, J. Chem. Phys. {\bf 45}, 3038 (1966)
\bibitem{plazek-bero} D. J. Plazek, C. A. Bero and I.-C. Chay, J. Non-Cryst. Solids {\bf 172-174}, 181 (1994)
\bibitem{mills} J. J. Mills, J. Noncryst. Solids {\bf 14}, 255 (1974)
\bibitem{schro} K. Schr\"oter, G. Wilde, R. Willnecker, M. Weiss, K. Samwer, and E. Donth, Eur. Phys. J. {\bf 5}, 1 (1998)
\bibitem{donth1} K. Schr\"oter and E. Donth, J. Chem. Phys. {\bf 113}, 9101 (2000)
\bibitem{donth2} K. Schr\"oter and E. Donth, J. Non-Cryst. Solids {\bf 307-310}, 270 (2002)
\bibitem{niss1} K. Niss, B. Jakobsen and N. B. Olsen, J. Chem. Phys.
{\bf 123}, 234510 (2005); B. Jakobsen, K. Niss and N. B. Olsen, J. Chem. Phys.
{\bf 123}, 234511 (2005)
\bibitem{maggi} C. Maggi, B. Jakobsen, T. Christensen, N. B. Olsen and J. C. Dyre, J. Phys. Chem. B {\bf 112}, 16320 (2008)
\bibitem{pcnew} C. Gainaru, T. Hecksher, N. B. Olsen, R. B\"ohmer, and J. C. Dyre, J. Chem. Phys. {\bf 137}, 064508 (2012)
\bibitem{tina} T. Hecksher, N. B. Olsen, K. A. Nelson, J. C. Dyre and T. Christensen, J. Chem. Phys. {\bf 138}, 12A543 (2013)
\bibitem{cavagna} A. Cavagna, Phys. Rep. {\bf 476}, 51 (2009)
\bibitem{bb} L. Berthier and G. Biroli, Rev. Mod. Phys. {\bf 83}, 587 (2011) 
\bibitem{tina2010} T. Hecksher, N. B. Olsen, K. Niss, and J. C. Dyre, J. Chem. Phys. {\bf 133}, 174514 (2010)
\bibitem{bo} B. Jakobsen, T. Hecksher, T. Christensen, N. B. Olsen, J. C. Dyre, and K. Niss, J. Chem. Phys. {\bf 136}, 
\bibitem{visc} U. Buchenau, Phys. Rev E {\bf 93}, 032608 (2016)
\bibitem{nonlin} U. Buchenau, arXiv:1512.00188
\bibitem{ferry} D. J. Ferry, "Viscoelastic properties of polymers", 3rd ed., John Wiley, New York 1980
\bibitem{hetero} R. Richert, J. Phys.: Condens. Matter {\bf 14}, R703 (2002)
\bibitem{eshelby} J. D. Eshelby, Proc. Roy. Soc. {\bf A241}, 376 (1957)
\bibitem{lunken} M. K\"ohler, P. Lunkenheimer, Y. Goncharov, R. Wehn, and A. Loidl, J. Non-Cryst. Solids {\bf 356}, 529 (2010)
\bibitem{ngai-rendell} K. L. Ngai and R. W. Rendell, Phys. Rev. B {\bf 41}, 754 (1990)
\end{thebibliography}
\end{document}